\documentclass[aoas,nameyear,seceqn,dvips]{arximspdf}
\usepackage{mathbh}
\usepackage{eurosym}
\usepackage{dcolumn}
\usepackage{graphicx}

\doi{10.1214/10-AOAS443}
\volume{5}
\issue{2B}
\pubyear{2011}
\firstpage{1632}
\lastpage{1656}

\makeatletter

\newproclaim{ass}{Assumption}

\newcolumntype{d}[1]{D{.}{.}{#1}}
\def\eqref#1{(\ref{#1})}

\makeatother

\begin{document}
\begin{frontmatter}

\title{Hierarchical Bayesian estimation of inequality measures with
nonrectangular censored survey data with an application to wealth
distribution of French households}
\runtitle{Estimation of wealth inequality}

\begin{aug}
\author{\fnms{Eric} \snm{Gautier}\corref{}\ead[label=e1]{eric.gautier@ensae-paristech.fr}
\ead[label=u1,url]{www.crest.fr/pageperso/eric.gautier/eric.gautier.htm}}
\runauthor{E. Gautier}
\affiliation{CREST (ENSAE)}
\address{ENSAE\\ 3 avenue Pierre Larousse\\ 92
245 Malakoff Cedex\\ France\\
\printead{e1}\\
\printead{u1}}
\end{aug}

\received{\smonth{5} \syear{2009}}
\revised{\smonth{11} \syear{2010}}

%
\begin{abstract}
We consider the estimation of wealth inequality measures with their
confidence interval, based on survey data with interval censoring.
We rely on a Bayesian hierarchical model. It consists of a~model
where, due to survey sampling and unit nonresponse, the summaries
of the wealth distribution of households are observed with error; a~%
mixture of multivariate models for the wealth components where
groups correspond to portfolios of assets; and a prior on the
parameters. A Gibbs sampler is used for numerical purposes to do the
inference. We apply this strategy to the French 2004 Wealth Survey.
In order to alleviate the nonresponse, the amounts were
systematically collected in the form of brackets. Matched
administrative data on the liability of the respondents for wealth
tax and response to overview questions are used to better localize
the wealth components. It implies nonrectangular multidimensional
censoring. The variance of the error term in the model for the
population inequality measures is obtained using linearization and
taking into account the complex sampling design and the various
weight adjustments.
\end{abstract}

%

\end{frontmatter}


\section{Introduction}\label{s0}
The estimation of wealth inequality measures for a gi\-ven finite
population (e.g., a country) is a difficult problem. A main
complicating issue is that wealth can be defined in different ways.
Data on wealth can be obtained from numerous sources---banks,
notaries (inheritances), tax declarations (e.g., wealth taxes) and
surveys among them---that may differ in their exact definitions.
Fundamentally, these sources are often limited to information on
particular elements on wealth, and so do not provide good
indications of total net worth (that is, the current value of all
marketable or fungible assets less the current value of total
liabilities or debts). The sources may also not be representative of
complete populations of interest; for instance, data on a tax
focused on high wealth brackets are inherently limited to just those
persons above the designated threshold.

Household surveys on wealth are a common way to collect data from
wider populations. American wealth surveys include the Survey of
Consumer Finance (SCF) and the wealth extensions of the Panel Study
of Income Dynamics (PSID). France's public office for statistics
and economic studies, INSEE, designs and administers the wealth
survey known as the Enqu\^{e}te Patrimoine (hereafter referred to as
EP). Though these surveys can usefully collect substantial amounts
of information, they are far from perfect as measures of wealth.
The personal or intrusive nature of wealth questions and their level
of detail subject them to potentially high nonresponse rates (due,
perhaps, to fear of theft or confusion between the data collector
and tax authorities). It has been observed in the SCF that
nonresponse is higher among the rich [\citet{Kenn}], for whom
answering the survey takes a~much longer time simply because assets
are more numerous. Wealth can also be inherently difficult to
discuss accurately---for instance, it is difficult to know the
``market value'' of one's personal or small business assets without
actually bringing them to market.

To ease collection of wealth information and to make the questions
easier and less intrusive to answer, it is now common to ask for
bracket information rather than specific amounts. In some surveys,
intervals may be the only responses; in others, displaying flash
cards and asking for responses within particular intervals may be
used as a remedy when a respondent is hesitant or unable to provide
a single amount. \citet{CG} and \citet{JS} discuss the conceptual
advantages and disadvantages of the collection of bracket data; the
use of categorical, interval data or the mixing of bracket and
point-specific data also raise analytical challenges.

This paper addresses the specific challenges in using survey data to
study wealth inequality: the extent to which wealth is unevenly
distributed across the population, such as a small share of people
holding a large share of the wealth in a population group.
Accordingly, one further complication of survey-based data on wealth
merits mention. Household surveys should adequately represent the
whole distribution of wealth, but the variance of
sample-survey-based estimates of wealth inequality can be reduced by
oversampling the wealthy. The major surveys can vary greatly in the
way they do this: the PSID is principally targeted at studying
lower-income populations and thus not well suited for wealth
inequality measures, while the SCF's dual-frame design includes a
list sample of households likely to be wealthy, using a
stratification based on variables from individual tax returns.

In this paper we utilize data from the 2004 administration of the
French EP, the design of which was developed to address these
methodological issues. The survey asks only for interval measures
for amounts of wealth; for some assets, the EP asked respondents to
choose categorical brackets from reference cards, but in others
respondents could specify their own bounds. The survey also mixed
questions on specific components of wealth with overview questions,
as a check on consistency. To estimate total net worth, information
from the overview questions, individual components and limited
matching to tax data (liability under a French wealth tax) can be
used to provide tighter estimates. The EP oversamples very wealthy
households via a stratification based on proxies of wealth. Because
of these features, the EP survey design is very complex; confidence
intervals are hard to obtain even in the ideal cases where tight
values of total net worth are observed for all sampled households
[see, e.g., \citet{SSW}]. The information on wealth that results
from the EP are a set of intricate domains, making it difficult or
impossible to directly calculate wealth inequality measures.

This paper develops a solution for estimating wealth inequality
based on a Bayesian hierarchical model. We begin in Sections
\ref{s1} and \ref{s2} by describing the data source---the 2004 EP
survey---in more detail, covering the survey design and the
comparison of EP results with other data sources. Section \ref{s3}
introduces the inequality indices and the design based procedure to
provide an interval estimate in the ideal case where there is perfect
response. Section \ref{s4} presents the hierarchical model. Section
\ref{s5} describes the multivariate domains used as an information set
for the posterior inference. Section \ref{s6} deals with the
specific approach to inference. Section \ref{s7} presents the Gibbs
sampler used for numerical purposes. Section \ref{s8} presents the
results for the 2004 EP. Section \ref{s9} concludes.

\section{The 2004 French Wealth Survey}\label{s1}
\subsection{General overview}
Administered approximately every 6 years since 1986, the EP has
become a critical reference on wealth in France. Unlike the
American surveys, response to the EP is mandatory rather than
voluntary. The EP provides information on wealth portfolios and the
distributions of a large number of assets of French households. It
also collects information on current and past employment, marital
history, income, transmissions, the modes of acquisition of the
principal residence, debts, credit, risk aversion, etc. EP data are
widely used by three key constituencies: by INSEE to establish the
national accounts on wealth and as input to the French
microsimulation model, by the French central bank (which partially
funds the survey collection), and by external researchers studying
wealth inequalities and dynamics.

\subsection{The sampling scheme, weighting and data collection}
The collection of the 2004 EP data took place from October 2003 to
January 2004. It is a~survey on households in their principal
residence. The sample design has two phases. The first phase is
common for all surveys on households in France, previous to the
renovated French census, and corresponds to sampling in two sampling
frames: the ``Master Sample'' (constructed from the 1999 census), and
a~sampling frame of real estate built after 1999. The Master Sample
is a~sampling frame of cities or groups of smaller neighboring towns
or districts for larger cities. It was obtained using a stratified
cluster sampling with two or three stages, depending on the stratum.
The 5 strata correspond to the following: (1) the rural, (2) urban
units with less
than 20,000 inhabitants, (3) between 20,000 and 100,000, (4) more
than 100,000 excluding Paris, and (5) Paris. The first phase of the
2004 EP corresponds, therefore, to a stratified three to four stage
sampling. In the first phase, 40,079 households were sampled. In
the second phase, 15,025 households were sampled according to a
stratified sampling with unequal probabilities. 10 strata were
chosen: 8 for principal residences at the time of the census, 1 for
other dwellings at the time of the census and 1 for real estate
built after 1999. Unequal probabilities were used to include a
priori more wealthy households. We present, in Table~\ref{Tasamp},
the proportions corresponding to the second phase oversampling.

\begin{table}
\centering
\caption{Second phase oversampling of principal residences}\label{Tasamp}
\begin{tabular*}{\tablewidth}{@{\extracolsep{4in minus 4in}}lcccc@{}}
\hline
& \textbf{Self-employed and} &  & \textbf{Retired} & \\
& \textbf{company owners} &\textbf{Executives} & \textbf{people} &  \textbf{Others}\\
\hline
Rich neighborhoods & 4 & 3\phantom{.0} & 3\phantom{.0} & 2 \\
Other neighborhoods & 2 & 1.5 & 1.5 & 1 \\
\hline
\end{tabular*}
\end{table}

The initial weights were modified because they implied an estimate
of 57.1\% of home owners at the time of census, while the true
percentage was 54.7\%. Among the sampled units, 13,154 dwellings
corresponded to principal residences and were kept. Eventually, due
to unit nonresponse, 9692 questionnaires remained. Sampling
weights were adjusted again to account for unit nonresponse, using
stratification and assuming a uniform nonresponse mechanism per
strata. The initial weights were divided by response rates per
strata. The unit nonresponse is traditionally modeled as a third
phase Poisson sampling and the new weights are usually treated as if
they were the true inverse of the inclusion probabilities: we
propose an alternative method in Section \ref{s9}. In order to
decrease the variance of the survey sampling estimators and to
account for the changes in the French population since the 1999
census, a calibration procedure was used [\citet{DS}]. More details
on the design, unit nonresponse adjustment and calibration are
available on the survey's
webpage.\footnote{\url{http://www.insee.fr/fr/themes/detail.asp?reg_id=0&ref_id=fd-patri04}.}

\subsection{The survey questionnaire}
The survey questionnaire comprised two parts of unequal
length. The first part was face-to-face interviews using
computer-assisted personal interviewing (CAPI), like for the SCF. A
second questionnaire on general attitudes and risk exposure was left with
the households, to be returned by mail in a prepaid envelope.

The CAPI questionnaire was organized as
follows: the first section gathered information on the people in the
household; the second section was concerned with holdings of assets
and liabilities; sections were then organized according to types of
assets, and amounts were collected in brackets; then data on income,
loans, donations, inheritance, debts and life
annuities was collected.

$\!$The section on financial wealth gathered information on every type
of~finan\-cial asset: checking accounts, saving accounts, CD accounts,
profit sharing, corporate savings plans, pension schemes,
participating insurances, stocks, bonds, etc. For the market value
of each asset, people were asked to choose a~bracket within asset
specific range cards. For example, in the case of checking accounts
and amounts in euros, the following system was used:
\begin{eqnarray*}
[0, 750),\qquad [750, 1500),\qquad [1500, 3000),\qquad [3 000, 7500),\qquad [7500, \infty).
\end{eqnarray*}
At the end of this section, an overview question was asked:

``Taking into account everything that you own, what is the value of
your entire financial wealth?''

The amount was collected within the following ranges:
\begin{eqnarray*}
&{}[0, 3000),\qquad [3000, 7500),\qquad [7500, 15{,}000),\qquad [15{,}000,
30{,}000), &
\\
&{}[30{,}000, 45{,}000),\qquad [45{,}000,
75{,}000),\qquad [75{,}000, 105{,}000),&
\\
&{}[105{,}000, 150{,}000),\qquad [150{,}000, 225{,}000),\qquad [225{,}000, 300{,}000), &
\\
&{}[300{,}000, 450{,}000), \qquad[450{,}000, \infty).&
\end{eqnarray*}
There were also overview questions for some blocks of assets.

The section on wealth in real estate gathered information on the
principal residence, holiday homes, pied-\`{a}-terres,
rentals and private parking lots.

The section on professional wealth gathered information on assets
and liabilities potentially related to the exercise of a profession.
There was a~distinction between those which are directly related to
a profit-generating occupation in the case of the self-employed or
company owners, and those which are not. In the first case, the
liabilities are loans and the assets are farmed lands, vineyards,
orchards, woods, other lands, buildings, machinery, equipment,
vehicles, livestock, stock, clientele, commercial/farming leases,
etc. In the second case, the assets are lands, buildings,
machinery, equipment, vehicles, livestock, stock, etc., which are not
used to generate profit. For all the amounts which are not related
to financial wealth, people were asked to provide a bracket with
limits that they could choose based on their evaluation.\looseness=-1

A specific question concerning total wealth was asked at the end of
the section gathering amounts:
\begin{quote}
``Suppose you sell everything, including durable goods, works of art,
private collections,
precious metals and jewelry, how much could you get for it?''
\end{quote}
The values of the last items were not collected in the previous
sections. Indeed, it could have been troublesome if the pollster
asked for such information and a robbery occurred after the visit.
The amount was collected within the same predefined system of
brackets as for the overview question on financial wealth. The
threshold for the higher and unbounded bracket is 450{,}000 \euro{}.
It was chosen well below the threshold of 720{,}000 \euro{} for the
liability for the ISF (Imp\^{o}t Sur la Fortune, a specific French
wealth tax) in order to mitigate the nonresponse rate.

In Table \ref{Tares} we compare 3 variables in terms of the type of
response that was obtained. Figures are percentages out of the
responding households, sample weights are not taken into account.
Point measures occur when the respondents provide their own limits
to the bracket and when these limits are equal. When we consider
wealth components at an aggregate level, with a sum of detailed
wealth components, as soon as one component is measured in interval,
the sum falls into some interval. We see in Table \ref{Tares} that
genuine item nonresponse is relatively low.

\begin{table}
\caption{Type of response for different variables}\label{Tares}
\begin{tabular*}{\tablewidth}{@{\extracolsep{4in minus 4in}}ld{2.1}d{3.1}d{3.1}@{}}
\hline
\textbf{Share of:} & \multicolumn{1}{c}{\textbf{Principal}} & \multicolumn{1}{c}{\textbf{Financial}} & \multicolumn{1}{c@{}}{\textbf{Total wealth}} \\
\textbf{(in percent, without weighting)} & \multicolumn{1}{c}{\textbf{residence}} & \multicolumn{1}{c}{\textbf{wealth}} & \multicolumn{1}{c@{}}{\textbf{(last question)}}
\\
\hline
Holdings & 55.7 & 100 & 100 \\
Point measures & 12.3 & 0 & 0 \\
Unbounded brackets & 2.8 & 0.7 & 7.5 \\
Bounded brackets & 76.6 & 94.4 & 86.7 \\
Item nonresponse & 8.3 & 4.0 & 4.8 \\
\hline
\end{tabular*}
\end{table}

\section{Quality of the data, comparison and matching with
administrative data}\label{s2}
Brackets for components and those involving several components
(overview questions on some groups of financial assets, the total
financial wealth and the total wealth) were not always coherent.
This enabled the detection of errors like confusion between Francs
and Euros or errors due to the difficulty in recall when summing
amounts. Consistency checks based on these overview questions were
used during the CAPI administration of the survey.

A fraction of the households surveyed in the 2004 EP have been
interviewed later by sociologists in order to learn how the survey
was perceived. It was mainly aimed to understand the households'
difficulties to talk about money and wealth. Overall, the households
felt a sense of civic responsibility to answer the questions. They
found it less confidential to answer questions about holdings than
questions about amounts. They seemed to know quite well their
wealth holdings and talked very easily about their principal
residence. The financial wealth was a more difficult topic. For
example, though the surveys asked for the current value of each
asset, many households answered the value initially invested and
found it difficult to take into account the appreciation or
depreciation when they had not cashed it or sold the asset. For
more information on these interviews see \citet{CGi} and the
references therein.

Concerning wealth holdings, we will make the assumption that the
information on holdings is always accurate. This is certainly only
partially true. However, questions on holdings are indeed less
indiscreet than questions on the values of the assets. Moreover,
the questionnaire was designed so that very early, right after the
collection of the information on the households members, questions
on holdings were asked without any reference to the amounts. In
this synthetic block, answering yes or no thus took the exact same
time. It is only later, once the full portfolio of wealth was known,
that questions on amounts were asked. It did not appear from the
testing of the questionnaire that there was bias on the holdings of
products on the bottom of the list. Comparison of the results on
holdings of financial assets in the EP with data provided by banks
(gathered by the French central bank) have proved, in the past, to
be very satisfactory. The publication of the results on holdings by
INSEE is judged satisfactory by the professionals that use it. What
occurred often, though, is people who declared in the first stage
that they hold a product but then refused to give a bracketed
amount.

There is another issue with the values of the components of wealth
which is related to the type of data that is collected. The last
question of the section on amounts which collects the total net
worth used a system of broad intervals, topcoded at a relatively low
value in order that the households do not suspect a tax
investigation and provide an answer to the question. Based solely
on this question, a billionaire is observationally equivalent to a
household whose total wealth is 450{,}001 \euro{}. Though in theory
oversampling more a priori wealthy people improves the accuracy of
estimators of inequality indices like the Gini; in practice, because
we collected less precise information on the wealthiest,
oversampling increased the number of households for which we
measured wealth inaccurately. Because it is important to have a good
picture of the wealth, especially for the wealthy who contribute
significantly to the inequality, it is useful to gather the most
adequate information on the total net worth and the wealth
components. This is why we not only use the last overview question
but use also aggregated wealth components.

We were also able to match the survey data with a file provided by
the French tax authority which gives the tax liability of the
surveyed households for the 2004 ISF, a specific tax, paid only by
wealthy households. Taxable wealth is very different from total net
worth we are interested in. Still, it is, as we will see, very
useful to anchor the values of the wealth components and provide for
each responding household a smaller multidimensional domain
containing the values of the aggregated wealth components.

\section{Inequality indices and survey sampling estimators}\label{s3}
\subsection{Inequality indices}
For the sake of completeness we present the three inequality
measures that we use: the Gini (based on the Lorentz curve), the Atkinson
family and the Theil.

The Lorentz curve plots the proportion of national wealth earned by
each given percentage of households, ordered from the poorest to the
richest. It is increasing and convex. Complete equality corresponds
to a straight 45 degree line through the origin. In this case the
poorest $x$\% of households possess $x$\% of the national wealth.
The greater the departure from this straight line, the higher the
concentration of wealth among a relatively small number of
households. The Gini index corresponds to twice the area between
the straight line of equal distribution and the Lorentz curve. The
closer it is to one, the higher the concentration. If we denote by
$t_k$ the (total) wealth of the household of index $k$ from 1 to
$N$, $N$ the total number of households in the French population,
$r(k)=\sum_{i=1}^N\mathbh{1}\{t_i\le t_k\}$ the rank of $t_k$,
$\mathbh{1}\{\cdot\}$ the indicator function and
$\overline{t}=\frac1N\sum_{k=1}^Nt_k$, the formula for the Gini is
\[
G=\frac{\sum_{k=1}^N(2r(k)-1)t_k}{N^2\overline{t}}-1.
\]

The inequality measures introduced in \citet{Atk} are
\begin{eqnarray*}
I=1-\frac{U^{-1}((1/N)\sum_{k=1}^NU(t_k)
)}{\overline{t}},
\end{eqnarray*}
where $U$ is a utility function which is increasing and concave and
the numerator is the equally distributed equivalent of total wealth
corresponding to the expected utility (or social welfare function).
They lie between zero and one. The closer they are to one, the more
unequal the distribution of wealth. Interpretation is easy: if
$I=0.9$, then we would need only 10\% of the national wealth to achieve
the same level of social welfare. Under the constant relative
inequality aversion assumption, which corresponds to the requirement
that $I$ is homogeneous of degree zero (i.e., invariant with respect
to proportional changes in wealth), the function $U$ is necessarily
among a~specific one parameter family of functions [\citet{Atk}].
Hence, we get the following family of inequality indices indexed by
$\varepsilon>0$:
\begin{eqnarray*}
A_{\varepsilon}&=&1-\Biggl(\frac1N
\sum_{k=1}^N\biggl(\frac{t_k}{\overline{t}}\biggr)^{1-\varepsilon
}\Biggr)^{1/{(1-\varepsilon)}}\qquad\mbox{if } \varepsilon\ne1,
\\
A_{1}&=&1-\Biggl(\prod_{k=1}^N\frac{t_k}{\overline{t}}\Biggr)^{1/N}.
\end{eqnarray*}
Because $\varepsilon$ is a measure of inequality aversion, higher
values of $\varepsilon$ lead to more weight being attached to transfers
at the lower end of the distribution.

The inequality measure introduced in \citet{Th}, derived from
entropy, is defined by
\[
T=\frac{1}{N}\sum_{k=1}^N\frac{t_k}{\overline{t}}\log\biggl(\frac{t_k}
{\overline{t}}\biggr).
\]
The Theil decomposability holds: in a
population consisting of several groups, inequality can be expressed
as the sum of within group inequality and between group inequality.
The first is the sum of the inequality levels of each group weighted
by the share of national wealth it receives. The second is the
inequality index computed on average values, where we replace each
individual wealth by the average wealth of each group. As shown in
\citet{F}, this property is characteristic of the Theil index among
inequality measures that: (1) satisfy the Pigou--Dalton transfer
principle (inequality increases under a transfer from the poor to
the rich); (2) are invariant under permutations of the individual
wealth; and (3) are homogeneous of degree zero.

\subsection{Design based point and interval estimates}
We present in the case of the Gini index, and if wealth components
were observed, classical survey sampling estimators to obtain
confidence intervals. Recall that in Section~\ref{s1} (see also
Section~\ref{s5}), for the most part, only brackets with
possibly unbounded upper and/or lower bounds are available. Thus, in
reality, wealth components are not observed. The formulas for the
estimators and the variance calculations presented below cannot be
applied. We present in Section~\ref{s4} a hierarchical Bayesian
model to deal with this missing data problem.

Given sampling weights $(w_k)_{k=1}^N$, a design-based
estimate of the Gini is
%
\begin{equation}\label{eGhat}
\hat{G}=\frac{\sum_{k\in S}(2\hat{r}(k)-1)w_kt_k}{\sum_{k\in
S}w_k\sum_{k\in S}w_kt_k}-1,
\end{equation}
where $S$ is the randomly drawn set of indices of sampled
households and $\hat{r}(k)=\sum_{j\in S}w_j\mathbh{1}\{t_j\le
t_k\}$ is the estimated rank of the wealth of the household
of index $k$.

Hereafter, we denote by $m$ the cardinal of $S$. In practice, a
normal approximation for the design-based estimate is usually used
in order to obtain interval estimates. Justification of the
asymptotic normality of quite general nonlinear estimators, such as
that of the Gini, in the case of stratified two-stage sampling is
given in \citet{Shao}. It is also proposed to use the jackknife to
obtain an estimate of the asymptotic variance. Asymptotics in
survey statistics assume that the finite population quantities
correspond to draws in a super-population. Besides the jackknife,
other methods can be used. In this article, we decided to proceed
as explained in \citet{DSS}. It is based on the following: (1)
using linearization, under fairly general assumptions, we can
approximate the variance of a complex statistic by the variance of a
Horvitz--Thompson type estimator where the observations are the
linearized variables; (2) the variance of the new estimator can be
decomposed into several separate variances to account for
stratification, multistages and multiphases sampling; and (3) each
variance is approximated, using analytic formulas for each simpler
sampling procedures [\citet{SSW}]. Unequal probability sampling of
fixed sample size was treated as a maximum entropy sampling. This
allows us to use variance approximations that use only the
first-order inclusion probabilities [see (2.3) in \citet{DSS} and
\citet{MT}] which are usually good approximations. Calibration
amounts to modifying the initial weights in such a way that the
estimated totals $\sum_{k\in S}w_kX_k^i$ for a set of variables
$X^i$ are in line with known totals. \citet{DS} show that this
improves the accuracy of the estimators. The whole variance
calculations for Horvitz--Thompson estimators, accounting for the
complex sampling scheme and calibration, can be obtained using the
POULPE software developed by INSEE [\citet{CDS}]. Linearization of
the estimators of the summary of the wealth distribution we are
interested in is easily obtained using the rules explained in
\citet{DSS} and \citet{DDHFM}.

\section{The hierarchical model}\label{s4}
We shall now use capital letters for random variables and lowercase
letters for realizations. We also use bold characters for vectors.

We now enter into a key part of the paper where we present a method
that allows us to adapt the methodology of Section \ref{s3}, which
requires precise measurements, to the case where only bracketed data
is available. Again, we restrict our attention for model (I) below
to the estimation of the Gini, but the methodology is used in Section
\ref{s8} for many summaries of the wealth distribution. We start
off from the approximation
\[
\hat{G}\approx G+\sqrt{\widehat{V(\hat{G})}}E,
\]
where $\hat{G}$ is an asymptotically normal design-based estimate of
the Gini, for example \eqref{eGhat}. The error term $E$ is a
standard centered Gaussian random variable. The variance estimate,
which can be computed as described in Section \ref{s3}, is denoted
by $\widehat{V(\hat{G})}$.

 Due to the measurement in a bracketed format, in practice,
$\hat{G}$ and $\widehat{V(\hat{G})}$ cannot be computed.
We rely on a three-stage model:
\begin{enumerate}
\item model (I) for the quantities of interest, here the Gini,
conditional on the wealth of the households in the sample
$(T_1,\ldots,T_m)=(t_1,\ldots,t_m)$,
\begin{eqnarray}\label{eI}
&&G=\hat{G}(t_1,\ldots,t_m)+\sqrt{\widehat{V(\hat
{G})}
(t_1,\ldots,t_m)}E,\nonumber
\\[-8pt]\\[-8pt]
&&\eqntext{E \mbox{ is a standard normal error term};}
\end{eqnarray}
\item model (DGP) for the wealth components of the sampled
households, the sum of which is equal to $T_k$ for household
$k$, conditional on the value of covariates and on parameters;
\item the prior distribution (P) of the parameters $\bolds{\Theta}$
of density $\pi(\theta)$.
\end{enumerate}
We make the following assumption.

\renewcommand{\theass}{(\Alph{ass})}
\setcounter{ass}{0}
\begin{ass}
  $E$ is independent of the distribution of $
(T_1,\ldots,T_m)$
 conditional on the covariates specified in the DGP.
\end{ass}

\subsection{Model (I)} In equation \eqref{eI} $G$ is random, though it
is assumed to have an unknown but fixed value in the finite
population of French households. Reverting the Gaussian
approximation to obtain interval estimates is classical in
statistics. Also, from the super-population argument (used for
asymptotics in survey statistics), it makes perfect sense to
consider the finite population quantities as random. Conditional on
$(T_1,\ldots,T_m)= (t_1,\ldots,t_m)$,
$\hat{G}(t_1,\ldots,t_m)$ and
$\widehat{V(\hat{G})}(t_1,\ldots,t_m)$ can be
computed using \eqref{eGhat} and the variance estimation procedure
of Section~\ref{s3}.

\subsection{Assumption \textup{(A)}} It corresponds to the missing at random
(MAR) [\citet{LR}] assumption for the selection of the sample and
the unit nonresponse. This holds for the first selection stage.
Indeed, the variables used in the unequal probability sampling of
dwellings in the Master Sample are available. Recall that sampling
from the sampling frame for new dwellings does not rely on unequal
probabilities. However, Assumption (A) requires that the unit
nonresponse mechanism is also missing at random, and, thus, that in
the DGP model we have included the adequate covariates allowing us
to ignore the nonresponse mechanism. We will see below that
Assumption (A) is also important to justify the use of the
conditional log-normal distribution.

\subsection{Model (DGP)} Households might or might not hold each
detailed component, and can have an arbitrary quantity of them
(e.g., checking accounts). We chose a model which is a mixture of
multivariate Gaussian linear models for the logarithms of the amount
of the held components of wealth and groups correspond to each
pattern of holdings. The DGP that we specify allows for
interdependence between the amounts of the wealth components held,
the type of holding portfolio and portfolio specific parameters.
This is very important and usually imputations, even multiple
imputations, are done independently between components which
potentially leads to biases and is not coherent with the portfolio
choice theory. The DGP that we specify is similar to that of
\citet{HLR}. However, here we shall allow for covariance matrices
that are specific for each pattern of holdings. Working at a more
aggregate level allows us to introduce more covariates. \citet{HLR}
work with 12 components, but do not include covariates. Introducing
covariates seems important both for the coverage of the interval
estimates (predictive performance) and for the treatment of the unit
nonresponse [see Assumption (A)].

\emph{Wealth categories}. Macro components have been chosen to be
as homogeneous as possible in order to have good explanatory
covariates. They are defined in terms of the blocks of the survey
questionnaire: (1) financial wealth, $W^1$; (2) the value of the
principal residence, $W^2$; (3) of real estate other than the
principal residence (including second homes for rentals or for
leisure and private parking lots), $W^3$; (4)~professional wealth,
$W^4$; and (5) the remainder, $W^5$. The remainder corresponds to
durable goods (including vehicules, etc.), works of art, private
collections, precious metal and jewelry. We grouped together all
professional wealth---whether or not it is used to generate profit---and rental/nonrental real estate properties to have bigger sample
sizes. From a history of wealth accumulation perspective, it would
be meaningful to differentiate between assets which yield returns,
like rentals, some professional wealth, financial assets and other
assets. Such a decomposition of wealth into 5 components implies,
in principle, $2^5$ patterns of holdings. For simplicity, we assume
that every household has some financial wealth (e.g., money in a
checking account) and some wealth in the form of remainder (e.g.,
durable goods). As a result, we are left with only $2^3=8$ different
groups. 59.36\% of households own a primary dwelling, 21.99\% other
real estate and 19.78\% professional wealth. Table \ref{Tarec}
gives the size of each of the eight groups.
We denote by $\mathbf{D}_k=(D_{k,l})_{l=1,\ldots,5}$ the binary vector
such that $D_{k,l}=\mathbh{1}\{W_k^l>0\}$ and define the map
$P$ which associates the index $i\in\{1,\ldots,8\}$ of the pattern
to each $\mathbf{D}_k$. The DGP for pattern $i$, that is, for $k$ such that
$P(\mathbf{D}_k)=i$, is
\begin{table}
\caption{Patterns of holdings}\label{Tarec}
\begin{tabular*}{\tablewidth}{@{\extracolsep{4in minus 4in}}lcccccccc@{}}
\hline
\textbf{Component/Group} & \textbf{1} & \textbf{2} & \textbf{3} & \textbf{4} & \textbf{5} & \textbf{6} & \textbf{7} & \textbf{8} \\
\hline
$W^1$ & $\surd$ & $\surd$ & $\surd$ & $\surd$ & $\surd$ & $\surd$
& $\surd$ & $\surd$ \\
$W^2$ & $\surd$ & $\surd$ & $\surd$ & & $\surd$ & & & \\
$W^3$ & $\surd$ & $\surd$ & & $\surd$ & & $\surd$ & & \\
$W^4$ & $\surd$ & & $\surd$ & $\surd$ & & & $\surd$ & \\
$W^5$ & $\surd$ & $\surd$ & $\surd$ & $\surd$ & $\surd$ & $\surd$
& $\surd$ & $\surd$ \\[3pt]
 Size & 658 & 984 & 837 & 147 & 3274 & 342 & 275 & 3175 \\
\hline
\end{tabular*}
\end{table}
%
\begin{equation}\label{eDGP}
\cases{\displaystyle
T_k=\sum_{l=1}^5s_k^lW_k^l,
\cr
\log(W_k^l)=\beta_{i,l}+\mathbf{x}_{k,l}\mathbf{b}_l+U_{k}^{l}&\quad \mbox{when}
$d_{k,l}=1$,
\cr
W_k^l=0 &\quad\mbox{when} $d_{k,l}=0$,
\cr
\mathbf{U_k}\rightsquigarrow\mathcal{N}(0,\Sigma_i),
}
\end{equation}
where $\mathbf{U_k}$ is a vector of size $p_i=\sum_{l=1}^5d_{k,l}$
gathering the components whe\-re~$W_k^l$ is nonzero. In order to use
product specific variables as covariates for the principal
residence, we model the value of the good. Thus, the share that the
household possesses is the multiplier $s_k^2$. In the other models,
for which the variables are sums of components collected in the
survey, we model the amount of the share that the household
possesses and use household specific variables only. Thus, for
$l\ne2$, $s_k^l=1$. We denote by $\mathbf{s}_k$ the stacked vector of
the $s_k^l$'s. We introduce fixed effects $\beta_{i,l}$ for the
type of portfolio. $\mathbf{x}_{k,l}$ includes a 1 to account for a
constant in the model. For identification, the~coeffi\-cient~%
$\beta_{1,l}$ is set to 0. This fixed effect allows us to account
for heterogeneity, and, since we do not allow $\mathbf{b}_l$ to depend
on $i$, permits a sufficiently large sample size for the estimation
of the regression coefficients for the logarithms. Other than these
group specific coefficients, the covariance matrices are also
allowed to depend on the type of portfolio allocation. Recall that~%
$W_k^l$ are unobservables and that only a domain that contains the
vector of held components is known. The parameters $\mathbf{b}_l$ and
$\Sigma_i$ are treated as unobservable random variables according to
the Bayesian paradigm [see model (P) below]. On the other hand, as
we mentioned previously, the variables $\mathbf{x}_{k,l}$, $d_{k,l}$
and $s_k^l$ are observables.

\begin{table}
\caption{Covariates for the DGP other than the type of portfolio}\label{Ta1}
\begin{tabular*}{\tablewidth}{@{\extracolsep{4in minus 4in}}lccccc@{}}
\hline
\textbf{Covariate/Component} & $\bolds{W^1}$ & $\bolds{W^2}$ & $\bolds{W^3}$ & $\bolds{W^4}$ & $\bolds{W^5}$ \\
\hline
\textit{Life cycle} & & & & & \\
\quad Single and childless & & $\surd$ & $\surd$ & $\surd$
& $\surd$ \\
\quad Age and age squared & & $\surd$ & $\surd$ & $\surd$
& $\surd$ \\
\quad Position in the life cycle & $\surd$ & & & & \\
\textit{Social and Education} & & & & & \\
\quad Social/professional characteristics & $\surd$ & $\surd
$ & $\surd$ & $\surd$ & $\surd$ \\
\quad Higher education degree & $\surd$ & $\surd$ & $\surd
$ & $\surd$ & $\surd$ \\
\textit{Income} & & & & & \\
\quad Level of the salary & $\surd$ & $\surd$ & $\surd$ &
$\surd$ & $\surd$ \\
\quad Social benefits received & $\surd$ & & & & \\
\quad Rent received &$\surd$ & $\surd$ & & $\surd$ & \\
\quad Other income received & $\surd$ & & $\surd$ & $\surd
$ & \\
\textit{Principal residence} & & & & & \\
\quad Location of the principal residence & $\surd$ & $\surd
$ & $\surd$ & & $\surd$ \\
\quad Surface and surface squared & & $\surd$ & & & \\
\quad Type of real estate & & $\surd$ & & & \\
\textit{History of wealth} & & & & & \\
\quad Donation received & $\surd$ & $\surd$ & & $\surd$ &
$\surd$ \\
\quad Donation given & $\surd$ & & & & \\
\quad Recent increase/decrease of wealth & $\surd$ & $\surd
$ & & $\surd$ & $\surd$ \\
\quad Type of wealth of the parents & $\surd$ & & $\surd$ &
$\surd$ & \\
\textit{Professional wealth} & & & & & \\
\quad Related to a profit generating occupation & & & &
$\surd$ & \\
\quad Firm owned & & & & $\surd$ & \\
\hline
\end{tabular*}
\end{table}

\emph{Covariates}. We summarize in Table \ref{Ta1} the covariates
introduced in the DGP.
Covariates include dummies (single and childless, social benefits
received, rent received, other income received, donations received,
donations given, recent increase/decrease in wealth, wealth carried
on business, firm owned), multinomials with $J$ alternatives
transformed into $J-1$ dummies (position in life cycle,
social/professional characteristics, higher education degree,
salary, location of the principal residence, type of real estate,
type of wealth of the parents) and continuous variables (age of the
principal adult, age squared, surface, surface squared). As usual,
introducing both the surface and the square of the surface is one
way to capture nonlinearities. Life cycle is a variable which
interacts age of the reference person and the type of family (single
person, childless couple, couple with one child, couple with two
children, couple with more than three children, single-parent
family, other). Selection of covariates was done marginal by
marginal where MLE is easy. We included variables (or proxies) from
the census that were used for oversampling (see Table \ref{Tasamp}),
unless they did not appear to be significant in the univariate
modeling of the wealth components. This is important because the
lognormal assumption could be justified in the general population
only. If the sampled households are endogeneously selected, then the
conditional distribution should not remain lognormal. We know that
the selection of the original sample (before unit nonresponse) is
exogeneous. This is also required for Assumption (A) to hold. Thus,
to avoid biases, we condition on the variables (or proxies) that
determine the selection process.

\subsection{Model (P)} We choose $\pi(\theta)$
proportional to
%
\begin{equation}\label{eP}
\prod_{i=1}^8\det(\Sigma_i)^{-{(p_i+1)/2}}.
\end{equation}
The vector of parameters $\bolds{\theta}$ in ${\mathbb{R}}^d$
corresponds to
the $(\beta_{i,l},\mathbf{b}_{l}')$'s and the matrices $\Sigma_i$
where, denoting by $\operatorname{dim}_l$ the dimension of any
$(\beta_{i,l},\mathbf{b}_{l}')$,
\[
d=\sum_{l=1}^5\operatorname{dim}_l+\frac12\sum_{k=2}^5k(k+1).
\]
The prior is a product of limits of normal/inverse-Wishart's
[\citet{LR}; \citet{S}], often called noninformative. The posterior, if the
data were observed, is a {\it bona-fide} normal/inverse-Wishart
probability distribution.

\subsection{The joint PDF}
The full joint pdf for the hierarchical model can be written with
usual notation
\[
f(G|{\mathbf
w}_1,\ldots, \mathbf{w}_m)\prod_{k=1}^mf(\mathbf{w}_k
|\theta,\mathbf{x}_k,\mathbf{d}_k,{\mathbf
s}_k)\pi(\theta).
\]
Recall that the vectors ${\mathbf
x}_k$, $\mathbf{d}_k$ and $\mathbf{s}_k$ are observables. However, the
vectors~$\mathbf{w}_k$ are not observed. We explain in Section
\ref{s5} that we are able to know, for each household, in what
domain $B_k$, $\mathbf{w}_k$ lies. The goal is now to carry on
inference on the posterior distribution of $G$ given the data: (1)
the vectors~$\mathbf{x}_k$, $\mathbf{d}_k$ and $\mathbf{s}_k$, and (2) the
domains $B_k$ containing the vectors $\mathbf{w}_k$; for
$k=1,\ldots,m$.

\section{Censoring and use of administrative data}\label{s5}
We explain in this section how we constructed the domains $B_k$
containing the vectors $\mathbf{w}_k$ for $k=1,\ldots,m$. First,
recall that we always know the status whether the household holds
the wealth component or not. We were easily able to build brackets
for the 5 macro components besides the remainder. The brackets for
financial wealth were obtained manipulating the overview question on
financial wealth and all the brackets for the held components of
financial wealth. Those for professional wealth were obtained
simply by summing the lower bounds and summing the upper bounds on
the values of the held components of professional wealth. For these
two components we do not have any point measures. We only have
brackets, possibly unbounded, or missing data. However, due to
equal upper and lower limits of the brackets, we do have $12.3$\%,
respectively $17.8$\%, of point measures for the value of the
principal residence and real estate other than the principal
residence. The bounds for the component $W_3$ were obtained by
summing lower bounds and by summing upper bounds. The information on
the total wealth, collected in the last question of the survey,
which includes the component $W_5$ that we call the remainder, allowed
to obtain upper and lower bounds on $W_5$. For this last component,
we do not have any point measures. The information on the remainder
is rather limited, especially for the top of the distribution of
wealth, but the liability for the ISF provides extra information on
the remainder (see below). One of the possible drawback of
aggregating components or collecting, for some components, brackets
among a predefined system exclusively, is the total absence of point
measures. In the absence of point measures, intervals are the main
information for identification and estimation. Also in the absence
of point measures, goodness-of-fit tests are unfortunately
impossible. The conditional lognormal distribution is commonly used
in the economic literature on wealth. We make such an assumption
for each marginal and allow for correlations of the error terms.
Alternative DGP could be formulated, for example, based on the Pareto
distribution. In any case, the rest of the methodology would be the
same with a different specification. Information in intervals are
used in Section \ref{s6} as an information set for the computation
of posterior means that are involved for the inference.

As we have seen, our data set was matched with restricted
data on the ISF. We are thus able to know which households pay the
ISF tax. The condition to be liable for the ISF is to have a
taxable wealth exceeding 720,000 \euro{}. We produced the following
upper and lower bounds on taxable wealth:
%
\begin{eqnarray}\label{eISFu}
&W_k^1+0.8s_k^2W_k^2+W_k^3+I_k\min(W_k^4,\mathit{ND}_{\max,k})+W_k^5-\mathit{DEBT}_k,&
\\\label{eISFl}
&W_k^1+0.8s_k^2W_k^2+W_k^3+\mathit{ND}_{\min,k}-\mathit{DEBT}_k ,&
\end{eqnarray}
where $\mathit{ND}_{\min,k}$ and $\mathit{ND}_{\max,k}$ are upper and lower bounds of
the nondeductible professional wealth obtained using the detailed
information, $I_k$ is a dummy variable indicating that some of the
professional wealth might not be deductible, and $\mathit{DEBT}_k$ is the
total of debts which are deductible. We assume that households
always subtract the deductible amounts. When a household pays the
tax, \eqref{eISFu} is greater than 720,000 \euro{}, while when it
does not pay the tax, \eqref{eISFl} is less than 720,000 \euro{}.
Only part of professional wealth is taxable. It is possible to
deduct the professional wealth related to a profit-generating
occupation if one's primary activity is self-employed, unless one
owns a share in a firm of less than $25\%$. It is possible to have
a~rebate of 20\% on the value of one's principal residence. Works
of art are not taxed and debts are deducted. It is possible to take
into account most of the characteristics of this tax and obtain
tight bounds. By chance, the few households that possessed a share
in a firm of less than $25\%$ gave a precise value of the firm. On
the other hand, it is impossible to distinguish works of art within
the remainder.

The final overview question on the total wealth, and
liability for the ISF, implies censoring domains which are subsets
of hyper-rectangles.

\section{The inference}\label{s6}
Suppose that the official statistician is asked to provide a single
value for each summary of the French wealth distribution. What is
the optimal answer? Specifying a loss function $l(\cdot,\star)$, it
is natural to minimize, among all answers $G^*$, the posterior
risk:
\begin{eqnarray}\label{ePR}
&&{\mathbb{E}}[l(G^*,G)|{\mathbf
W}_1\in B_1, \ldots,\mathbf{W}_m\in B_m,\nonumber
\\[-8pt]\\[-8pt]
&&\qquad\hspace*{32pt}{\mathbf
x}_1,\ldots,\mathbf{x}_m,\mathbf{d}_1,
\ldots,\mathbf{d}_m,\mathbf{s}_1,\ldots,\mathbf{s}_m ],\nonumber
\end{eqnarray}
where $G$ is given by the hierarchy of models from Section \ref{s4}.
It is classical that if a quadratic loss function is chosen, then
the optimal answer from a~risk minimization perspective is given by
the posterior mean
\begin{eqnarray}\label{ePred}
&&\overline{G}={\mathbb{E}}[G|{\mathbf
W}_1\in B_1,\ldots, \mathbf{W}_m\in B_m,\nonumber
\\[-8pt]\\[-8pt]
&&\hspace*{23pt}\qquad\mathbf{x}_1,\ldots,\mathbf{x}_m,\mathbf{d}_1,
\ldots,\mathbf{d}_m,\mathbf{s}_1,\ldots,\mathbf{s}_m].\nonumber
\end{eqnarray}
An interval estimate with confidence $1-\alpha$ can be obtained
finding $l\le u$ such that
\begin{eqnarray}\label{eCI}
&&{\mathbb{P}}(l\le G\le u|{\mathbf
W}_1\in B_1, \ldots,\mathbf{W}_m\in B_m,\nonumber
\\[-8pt]\\[-8pt]
&&\hspace*{37pt}\qquad{\mathbf
x}_1,\ldots,\mathbf{x}_m,\mathbf{d}_1,
\ldots,\mathbf{d}_m,\mathbf{s}_1,\ldots,\mathbf{s}_m )\ge
1-\alpha.\nonumber
\end{eqnarray}
Various types of such intervals are possible, including, for example,
HPD regions. One natural goal is to minimize the length of the
interval. Such interval estimates take into account both the usual
uncertainty related to sampling (sampling, unit nonresponse and
improvement of the accuracy due to calibration), and the uncertainty
due to the imperfect wealth measurement.\looseness=-1

\section{Monte Carlo Markov chain approximation}\label{s7}
According to Section \ref{s6}, inference relies on the evaluation of
integrals [\eqref{ePred} and \eqref{eCI}]. We use a Gibbs sampler
to simulate a path of a Markov chain $({\mathbf
v}_n)_{n\in\mathbb{N}}$ having as invariant probability $\mu$:
the joint posterior and posterior predictive and distribution of the
random disturbance $E$. Here, the $\mathbf{v}_n$'s could be
interpreted as scenarios~of
\[
\mathbf{V}=(\bolds{\Theta}',{\mathbf
W}_1',\ldots,\mathbf{W}_m',E)'.
\]
Limit theorems for the Gibbs
sampler can be found in \citet{RC}. Also, as in \citet{RP}, we can
prove uniform exponential $\mathrm{L}^1$ ergodicity by minorizing the
transition kernel. This follows from the fact that we introduced
upper bounds for the a priori unbounded amounts. Thus, convergence
of the distribution of the marginals of the Markov chain to the
target joint posterior and posterior predictive and distribution of
$E$ ($E$ is always independent of the rest of the components) should
be fast. The ergodic theorem yields approximations of the form
%
\begin{equation}\label{eergodic}
{\mathbb{E}}_{\mu} [g(\mathbf{V})]\approx
\frac
{1}{T-B}\sum_{n=B}^{T}g({\mathbf
V}_n)
\end{equation}
for some integer $B$ (burn-in) and large $T$. The Gibbs sampler is
a classical tool for simulation in truncated multivariate normals
[\citet{R95}] and in Bayesian statistics [\citet{RC}; \citet{MR}], including
in the multiple imputation literature [\citet{LR}; \citet{S}]. For the sake
of completeness, let us present the algorithm briefly. The Gibbs
sampler relies on a block decomposition of the coordinates of the
state space. These blocks are numbered according to a~specific
order. Starting from an initial value $\mathbf{v}_0$, the Gibbs sampler
simulates a~path from a Markov chain $({\mathbf
v}_n)_{n\ge0}$. Given $\mathbf{v}_n$, a vector $\mathbf{V}_{n+1}$
decomposed in the above system of blocks is simulated by iteratively
updating the blocks, and sampling from the distribution of the
block, conditional on the values at stage $n$ of the future blocks,
and the value at stage $n+1$ of the previously updated blocks. Here
$\mathbf{V}_n$ corresponds to
\[
(\bolds{\Theta}',{\mathbf
W}_1',\ldots,\mathbf{W}_m',E)'.
\]
The sequence is such that we
start by updating the $\mathbf{b}_l$'s, followed by the covariance
matrices, then one by one by the wealth components for each
household, and finish with the error term in model (I). It is
enough for the initiation of the algorithm to specify initial
conditions for the following: (1) the values of the held wealth
components of each
household in the sample, and (2) for the covariance matrices for
each group. We took as initial conditions for covariance matrices,
diagonal matrices, with diagonal terms being the estimated variances
of the error terms in the marginal models obtained by MLE. More
precisely, manipulations of the likelihood times prior imply the
sequence of simulations detailed below. We denote by ${\mathbf
b}=(\mathbf{b}_1',\mathbf{b}_2',\mathbf{b}_3',\mathbf{b}_4',{\mathbf
b}_5')'$, by $\mathbf{x}_k$ and $\mathbf{y}_k$ the matrices of size
$p_{P(\mathbf{d}_k)}\times\sum_{l=1}^5\operatorname{dim}_l$ and $p_{P({\mathbf
d}_k)}\times1$ extracted respectively from
\[
\pmatrix{
\mathbf{x}_{k,1} & 0\cdots0 & 0\cdots0 & 0\cdots0 & 0\cdots0 \cr
0\cdots0 & \mathbf{x}_{k,2} & 0\cdots0 & 0\cdots0 & 0\cdots0\cr
0\cdots0 & 0\cdots0 & \mathbf{x}_{k,3}& 0\cdots0 & 0\cdots0\cr
0\cdots0 & 0\cdots0 & 0\cdots0 & \mathbf{x}_{k,4} & 0\cdots0\cr
0\cdots0 & 0\cdots0 & 0\cdots0 & 0\cdots0 & \mathbf{x}_{k,5}
}
\quad\mbox{and}\quad
\pmatrix{
\log w_{k,1}\cr
\log w_{k,2}\cr
\log w_{k,3}\cr
\log w_{k,4}\cr
\log w_{k,5}
},
\]
where we only maintain the rows of index $l$ such that $d_{k,l}=1$.
At stage $n+1$, given the covariance matrices, values of the wealth
components and error term $E$ at stage $n$, we start by drawing
$\mathbf{b}_{n+1}$ in the multivariate normal
$\mathcal{N}(\hat\mathbf{b},\Sigma_\mathbf{b})$, where
\[
\cases{\displaystyle
\Sigma_\mathbf{b} = \Biggl(\sum_{i=1}^8
\sum_{k:P(\mathbf{d}_k)=i}\mathbf{x}_k'\Sigma_{i,n}^{-1}{\mathbf
x}_k\Biggr)^{-1},
\cr
\displaystyle \hat\mathbf{b} = \Sigma_\mathbf{b}^{-1}\Biggl(\sum_{i=1}^8
\sum_{k:P(\mathbf{d}_k)=i}\mathbf{x}_k'\Sigma_{i,n}^{-1}{\mathbf
y}_k\Biggr).
}
\]
We then sample the inverse of the covariance matrices independently.
For wealth pattern $i$ we draw $\Sigma_{i,n+1}^{-1}$ in the Wishart
distribution $\mathcal{W}_{p_i}(m_i,V)$, where the degree
of freedom $m_i$ is the sample size of the wealth pattern $i$ and
the scale matrix is
\[
V=\sum_{k:P(\mathbf{d}_k)=i}({\mathbf
y}_k-\mathbf{x}_k\mathbf{b})'(\mathbf{y}_k-\mathbf{x}_k{\mathbf
b}).
\]
We then update the wealth components for all the
households in the sample. We split each vector $\mathbf{W}_k$ in
blocks of size one. This uses the classical conditioning in the
multivariate normal random variate and allows us to simulate the
wealth components in univariate truncated normals [see, e.g.,
\citet{R95} for efficient algorithms]. The intervals of truncation
for the current variable at each stage of the sequence are updated,
taking into account the previously simulated components for the same
household, and the various
inequalities discussed in Section \ref{s5}.

We finally sample an independent error term $E_{n+1}$.

The integrals \eqref{ePred} and \eqref{eCI} which are used in this
article for inference are of the form ${\mathbb{E}}_{\mu}[g({\mathbf
V})]$, where
$g(\mathbf{V})$ is either $G$ or $\mathbh{1}_{G\in[l,u]}$ and $G$ is given
by the hierarchy of models from Section \ref{s4}. We therefore use
approximations of the form~\eqref{eergodic}. Here, for each $n$,
each $G_n=g(\mathbf{v}_n)$ is obtained from $\mathbf{v}_n$, computing the
total wealth $(t_1^n,\ldots,t_m^n)$ for each household in
the sample and using \eqref{eI} with the error random disturbance
$e_n$ and
$\widehat{V(\hat{G})}(t_1^n,\ldots,t_m^n)$
computed as explained in Section \ref{s3}. If we are interested in a
different statistic, we simply replace in \eqref{eI} the estimate of
the Gini coefficient $\hat{G}$ and of its variance
$\widehat{V(\hat{G})}$, by the corresponding survey
sampling estimators. This could be done with the same sample path of
the Gibbs sampler. Note that, concerning the interval estimation,
the above MCMC method is not optimal to evaluate quantiles and the
procedure requires very large $T$. For this reason, we chose to
present, in Section \ref{s8}, 90\% posterior regions.

The values $\mathbf{v}_n$ can be interpreted as multiple imputations.
None are in the target distribution since there is only convergence
to the invariant probability. We have seen in Section \ref{s6}
that an optimal estimation (with respect to a quadratic loss
function) is given by the posterior mean. Thus, simple random
imputation which corresponds to producing one random scenario for~%
$G$ is nonoptimal, as the risk of producing such a value is higher.
Moreover, it does not allow to obtain interval estimates. If
$g(\mathbf{V})$ is nonlinear in the wealth components, then the
prediction of individual wealth is not a~proper imputation procedure
even for point estimation. It does not yield a prediction of
$g(\mathbf{V})$. This is the case for all the summaries of the wealth
distribution given in Section \ref{s8} besides the mean.

\begin{figure}[b]

\includegraphics{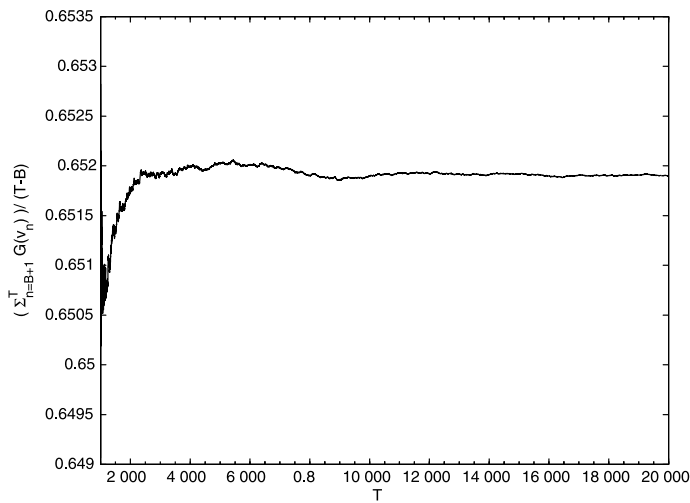}

\caption{Convergence of empirical averages of the Gini, $B=1 000$.}\label{fig1}
\end{figure}

\section{Presentation of the results}\label{s8}
\subsection{Results with the described DGP}\label{s81}
We ran a Gibbs sampler with $T=20{,}000$ and $B=1000$. In order
to diagnose convergence, we plotted the convergence of the empirical
averages required for the inference (see, e.g., Figure \ref{fig1}). As expected, due to exponential ergodicity, convergence
occurs very quickly. For such values of $T$ and $B$, burn-in only
changes the very last decimals.
For simplicity, for such plots, we used rough design-based variance
calculations based on linearization, but approximating the complex
sampling design. It is only below that we use the full procedure
explained in Section \ref{s3}. Since the computations in the POULPE
software are extensive, we take a larger value for $B$. We do not
feel that this is troublesome. Indeed, large $T$ is important for
convergence of the marginals of the Gibbs sampler to the invariant
probability. Once convergence is satisfactory, we compute the sample
analogues \eqref{eergodic}, starting close to the steady
state.

\begin{figure}

\includegraphics{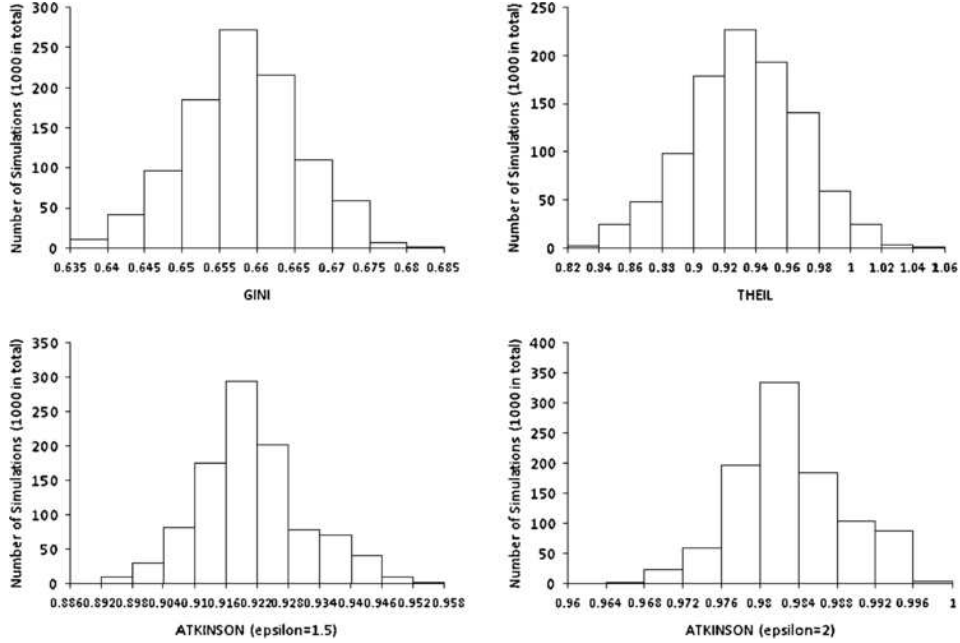}

\caption{Posterior distribution of the Gini, Theil and
Atkinson indices, full 5 components model, $T=20{,}000$ and $B=19{,}000$.}\label{fig2}
\end{figure}

In Table \ref{Ta2} we give posterior predictions and
confidence regions and in Figure \ref{fig2} we give histograms for
posterior distributions of summaries of the French wealth distribution.

\begin{table}
\caption{Posterior predictions and $90\%$ symmetric posterior
regions, \protect\eqref{eergodic} is used with $B=19{,}000$ and $T=20{,}000$}\label{Ta2}
\begin{tabular*}{\tablewidth}{@{\extracolsep{4in minus 4in}}lccc@{}}
\hline
\textbf{Summary of the distribution} & \textbf{Lower bound} & \textbf{Prediction} & \textbf{Upper bound}\\
\hline
Mean (\euro{}) & 202,600 & 211,200 & 218,800\\
Median (\euro{}) & 108,800 & 112,500 & 116,600\\
P99 (\euro{}) & 1,507,000 & 1,658,000 & 1,815,000\\
P95 (\euro{}) & 671,900 & 713,300 & 748,400\\
P90 (\euro{}) & 425,800 & 438,500 & 450,000\\
Q3 (\euro{}) & 228,300 & 234,000 & 239,600\\
Q1 (\euro{}) & 16,000 & 17,200 & 18,500\\
P10 (\euro{}) & 3324 & 3900 & 4459\\
P95/D5 & 5.97 & 6.33 & 6.64\\
P99/D5 & 13.30 & 14.71 & 16.17\\
Q3/Q1 & 12.71 & 13.72 & 14.50\\
D9/D1 & 94 & 111.2 & 126.4\\
D9/D5 & 3.75 & 3.89 & 4.02\\
Gini & 0.644 & 0.657 & 0.669\\
Theil & 0.870 & 0.930 & 0.984\\
Atkinson ($\varepsilon=1.5$) & 0.904 & 0.921 & 0.940\\
Atkinson ($\varepsilon=2$) & 0.974 & 0.983 & 0.993\\
\hline
\end{tabular*}
\end{table}

\subsection{Stability of the results regarding the aggregation
of wealth components}\label{s82}
 To study the relative stability of
the results regarding the aggregation of wealth components, we
present an alternative DGP model with fewer wealth components and
thus fewer wealth categories.

Suppose we decide to group together the values of the share held of
the principal residence and of the holdings in other real estate.
We now work with the following components: (1) financial wealth,
$\tilde{W}^1$; (2) wealth in real estate, $\tilde{W}^2$; (3) the
professional wealth, $\tilde{W}^3$; and (4) the remainder,
$\tilde{W}^4$. Table \ref{Tarec2} gives details about the size of
each of the $4=2^2$ groups.
The new wealth component is homogeneous in the sense that it is
investment in real estate. The choice is slightly less justifiable
from a wealth accumulation perspective, as principal residence and
other real estate are usually acquired one after the other. Also, the
second can yield returns. As a result, it is also possible to argue
that it is of a similar nature as some of the financial wealth. The
lower and upper bounds for this new aggregated component were
obtained by summing up respectively the lower bounds and upper
bounds of $s_2W_2$ and $W_3$. As a result, we only have for the new
component $11.9$\% of point measures. For all the other components
we do not have any point measures. We were no longer able to use
variables on the principal residence as covariates. For example, it
makes little sense to use the surface of the principal residence to
predict the value of the total share in real estate. In this case,
liability for the ISF is more difficult to exploit, as one is
allowed to have a rebate of 20\% on the value of one's principal
residence. We used rougher upper and lower bounds of taxable wealth
%
\begin{table}[b]
\tablewidth=300pt
\caption{Patterns of holdings}\label{Tarec2}
\begin{tabular*}{300pt}{@{\extracolsep{4in minus 4in}}lcccc@{}}
\hline
\textbf{{Component/Group}} & \textbf{1} & \textbf{2} & \textbf{3} & \textbf{4} \\
\hline
$\tilde{W}^1$ & $\surd$ & $\surd$ & $\surd$ & $\surd$ \\
$\tilde{W}^2$ & $\surd$ & $\surd$ & & \\
$\tilde{W}^3$ & $\surd$ & & $\surd$ & \\
$\tilde{W}^4$ & $\surd$ & $\surd$ & $\surd$ & $\surd$ \\[3pt]
{Size} & 1642 & 4600 & 275 & 3175 \\
\hline
\end{tabular*}
\end{table}
%
\begin{eqnarray}\label{eISFu2}
&\tilde{W}_k^1+\tilde{W}_k^2+\mathit{ND}_k\min(\tilde{W}_k^3,\mathit{NDED}_{\max,k})
+\tilde{W}_k^4-\mathit{DEBT}_k,&
\\\label{eISFl2}
&\tilde{W}_k^1+0.8\tilde{W}_k^2+\mathit{NDED}_{\min,k}-\mathit{DEBT}_k.&
\end{eqnarray}
When a household pays the tax, \eqref{eISFu2} is greater than 720,000 \euro{},
while when it does not pay the tax, \eqref{eISFl2} is
less than 720,000 \euro{}. In Table \ref{Ta22} we give posterior
predictions and confidence regions with the three-stage model with
this new DGP model. The interval estimates use calculations of the
asymptotic variances of the survey sampling estimators based on the
procedure presented in Section \ref{s3}.
This 4~components DGP yields results which are highly comparable to
those obtained for the 5 components DGP studied previously.

\begin{table}
\caption{Posterior predictions and $90\%$ symmetric posterior
regions ($T=20{,}000$, $B=19{,}000$)}\label{Ta22}
\begin{tabular*}{\tablewidth}{@{\extracolsep{4in minus 4in}}lccc@{}}
\hline
\textbf{Summary of the distribution} & \textbf{Lower bound} & \textbf{Prediction} & \textbf{Upper bound}\\
\hline
Mean (\euro{}) & 203,100 & 211,300 & 219,100\\
Median (\euro{}) & 108,700 & 112,600 & 116,400\\
P99 (\euro{}) & 1,498,000 & 1,661,200 & 1,822,300\\
P95 (\euro{}) & 673,100 & 714,000 & 749,700\\
P90 (\euro{}) & 426,300 & 438,800 & 451,000\\
Q3 (\euro{}) & 228,800 & 234,100 & 239,900\\
Q1 (\euro{}) & 16,020 & 17,210 & 18,470\\
P10 (\euro{}) & 3313 & 3914 & 4506\\
P95/D5 & 5.98 & 6.34 & 6.67\\
P99/D5 & 13.19 & 14.74 & 16.33\\
Q3/Q1 & 12.67 & 13.59 & 14.51\\
D9/D1 & 94.4 & 111.6 & 128.7\\
D9/D5 & 3.76 & 3.89 & 4.03\\
Gini & 0.644 & 0.658 & 0.670\\
Theil & 0.872 & 0.931 & 0.989\\
Atkinson ($\varepsilon=1.5$) & 0.904 & 0.921 & 0.940\\
Atkinson ($\varepsilon=2$) & 0.974 & 0.983 & 0.993\\
\hline
\end{tabular*}
\end{table}

\section{Concluding discussion}\label{s9}
In order to analyze the French wealth distribution based on the 2004
EP, we proposed a Bayesian hierarchical modeling. We produced point
and interval estimates of summaries of a finite population
distribution under random sampling, and in the presence of
generalized nonrectangular censoring. The approach is flexible, as
we can compute any possible such summaries (quantiles, inequality
indices, etc.), and is particularly useful when the summaries are
nonlinear in the input distribution. Unlike the original Bayesian
multiple imputation, we do not rely on proper---that is, independent---Bayesian
multiple imputations [\citet{LR}; \citet{S}], which could be
computationally intensive to obtain, nor rely on approximate
formulas to combine multiple imputations. Usually official
statisticians do not like to rely on models for the DGP. This does
not seem feasible in the presence of interval censored data and when
the sample survey estimator is ``nonlinear'' in the respondent's
wealth. It was, however, possible to take into account the complexity
of the sample design, auxiliary information on totals through
calibration, etc., using model (I). It is also possible to adopt a
model-based approach and to simulate the wealth for the nonsampled
households, but then the design features are not taken into account.
As noted in Section \ref{s1}, unit nonresponse was modeled as an
extra phase, resulting in estimated weights. As it is usually done
in practice, they were treated as the true inverse of the inclusion
probabilities. Interval estimates are thus slightly optimistic. One
way to deal with this problem is to treat the true weights as
observed with error and add an extra model in the hierarchy of
models. It implies to augment the state space of the Gibbs sampler
presented in Section \ref{s7}. We could also include uncertainty in
the model choice, including, for example, the possibility of a Pareto
distribution, with an additional model in the hierarchy and prior
weights on each model in competition. Indeed, distributional
assumptions made for the DGP are crucial especially for the
wealthiest. Finally, Assumption (A), made here for the unit
nonresponse, is a strong assumption that is made in most of the
literature on missing data in surveys. It is possible to relax this
assumption via strong parametric assumptions [\citet{Gau}]. These
extensions of the methodology proposed in this article could be
studied, for example, in a~simpler setting.

We favored objectivity and tried to impose the minimum possible
structure. For this reason, we used noninformative priors and did
not impose any structure on the covariance matrices in the DGP
model. A common practice is to assume diagonal covariance matrices
for the residuals. This is the case when imputations, possibly
multiple imputations, are done independently for each wealth
component. This is very questionable, as it is not coherent with the
portfolio choice theory. We feel that it imposes too much
structure. The cost for this objectivity is relatively large
interval estimates. We feel, though, that it is important for a
national statistical office to be as objective as possible.
Specification of the DGP components was taken to be the most
classical lognormal one. We traded off the number of parameters for
posterior regions with reasonable coverage. The model for the
multivariate DGP has a reasonably small number of components and
covariates for groups of small sample size. The components form
homogeneous blocks in terms of population and wealth accumulation
history. Observed heterogeneity is introduced through fixed effects
and covariates, unobserved heterogeneity through correlations of
error terms with group specific covariance matrices.

It is always useful to gather information from sources exterior to
the survey. This is difficult when one is using other survey data,
due to different concepts, different selection mechanisms,
especially because of unit nonresponse and the different perception
of surveys and different dates. Here we were able to use matched
administrative data for the same year to better localize the
interval censored wealth components.

Further improvement could be made for the measurement of wealth with
a sampling scheme designed explicitly for the study of the wealth
inequality. Because of its list sample, the SCF is probably better
designed for such studies. One possibility studied for the EP is to
draw households based on the wealth and property taxes (note that
the notion of household based on principal residences is different
from the one used for tax purposes), but it raises issues concerning
tax secrecy. In any case, there are limits to a~better sampling
design: confidentiality, the relative coarser information for the
wealthiest due to the collection of brackets, the general use of the
data; as well as limits inherent to social statistics: nonresponse,
biased responses, errors in recall for overview questions,
misunderstanding, etc.

\section*{Acknowledgments} The author thanks Christian Robert for his
guidance on MCMC and former colleagues at INSEE (including Marie
Cordier, C\'{e}dric Houdr\'{e}, Dominique Place and Daniel Verger), and
Yale for stimulating discussions. The author is also grateful to the
Editor Stephen Fienberg and the Associate Editor for useful
comments.

\printaddresses

\end{document}